\documentclass[fleqn,twoside]{article}
\usepackage{espcrc2}

\usepackage{graphicx}
\usepackage[figuresright]{rotating}

\newcommand{\AmS}{{\protect\the\textfont2
  A\kern-.1667em\lower.5ex\hbox{M}\kern-.125emS}}

\title{Updated Z-Burst Neutrinos at Horizons.
  }

\author{D. Fargion\address[MCSD]{Physics Department and INFN, \\
        Universit\`{a} La Sapienza, Roma, Ple. A. Moro 2, 00185, Italy.}, P. Oliva.%
        \thanks{The authors thank Dr. O.Lanciano and A. Colaiuda for useful discussions}}

\begin{document}

\begin{abstract}
Recent homogeneous and isotropic  maps of UHECR, suggest an
isotropic cosmic origin almost uncorrelated to nearby Local
Universe prescribed by GZK (tens Mpc) cut-off. Z-Burst model,
based on UHE neutrino resonant scattering on light relic ones in
nearby Hot neutrino Dark Halo, may overcome the absence of such a
local  imprint and explain the recent correlation with BL-Lac at
distances of a few hundred Mpc. Z-Burst multiple imprint, due to
very possible lightest non-degenerated neutrino masses, may inject
energy and modulate UHECR ZeV edge spectra. The Z-burst (and GZK)
ultra high energy neutrinos (ZeV and EeV band) may  also shine, by
UHE neutrinos mass state mixing, and rise in corresponding UHE Tau
neutrino flavor, whose charged current tau production and its
decay in flight, maybe the source of UHE showering on Earth. The
Radius and the atmosphere of our planet constrains the $\tau$
maximal distance and energy to make a shower. These
\textit{terrestrial} tau energies  are near GZK ($10^{19}$ eV).
Higher distances and energies are available in bigger planets (up
to ($6\cdot10^{19}$ eV)); eventual solar atmosphere horizons may
amplify the UHE tau flight allowing tau showering at ZeV energies
($3.5\cdot10^{20}$ eV), offering a novel way to reveal the
expected Z-Burst extreme neutrino fluxes. \vspace{1pc}
\end{abstract}
\maketitle
\section{UHECR Isotropy and Spectra.}
\begin{figure}[t]
\includegraphics[width=70mm]{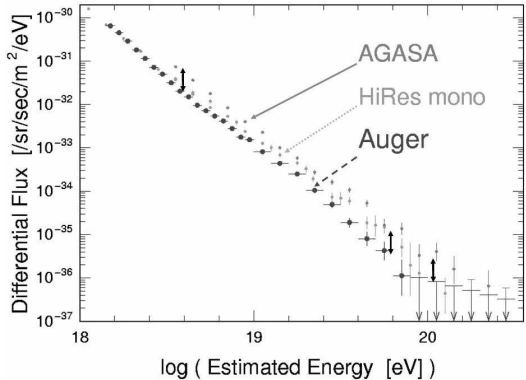}
\caption{\footnotesize{Last Auger UHECR  data versus HIRES and
AGASA ones; it is remarkable how the disagreement between the
spectra occurs already at few EeV energy band up to the most
energetic edges. The arrows shows the discrepancy all along the
UHECR spectra among the AGASA versus HIRES-AUGER experiments
\cite{Cronin2004}. }} \label{fig1}
\end{figure}
\begin{figure}[t]
\includegraphics[width=70mm]{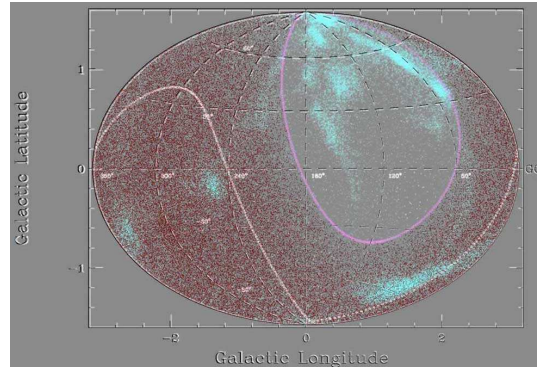}
\caption{\footnotesize{Last Auger UHECR data at EeV energy (71147
events) from AUGER and overimpress the expected local mass (GZK
volume) imprint, in negative colors. The absence of any viewable
galactic correlation or Local Group, at EeV energy (contrary to
AGASA and SUGAR claim) might be indebt to a galactic magnetic
field scrambling or to an overall dominance of extragalactic EeV
UHECR; nevertheless the AUGER dismiss of any rising UHE EeV
neutron astronomy in our Galaxy has been a surprise.}}
\label{fig2}
\end{figure}
\begin{figure}[t]
\includegraphics[width=70mm]{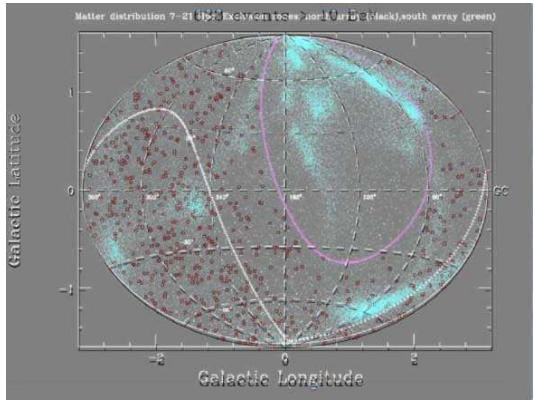}
\caption{\footnotesize{As above Last Auger UHECR data at ten EeV
energy (633 events) from AUGER and overimpress the expected local
mass (GZK volume) imprint, in negative colors. The absence of any
correlation, (as for AGASA and HIRES records) confirm a surprising
homogeneity and isotropy possibly of cosmic origin.}} \label{fig3}
\end{figure}
\begin{figure}[t]
\includegraphics[width=70mm]{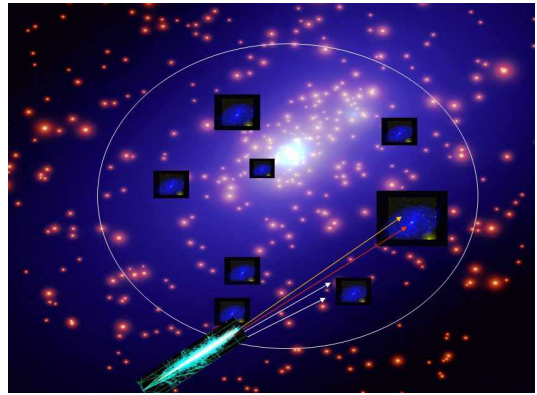}
\caption{\footnotesize{Z-Burst model is based on UHE ZeV neutrino,
ejected by far cosmic BL-Lac, AGN sources, scattering onto relic
light ones spread in Huge Galactic Group Hot dark Halos. The halo
radius maybe as large as GZK one and the Z decay might shower
leading, among tens of pions, and gamma traces, to UHE neutral
nucleons (neutron and anti-neutrons) or proton, anti-protons whose
propagation and hitting the Earth arise as UHECR.}} \label{fig4}
\end{figure}
\begin{figure}[t]
\includegraphics[width=65mm]{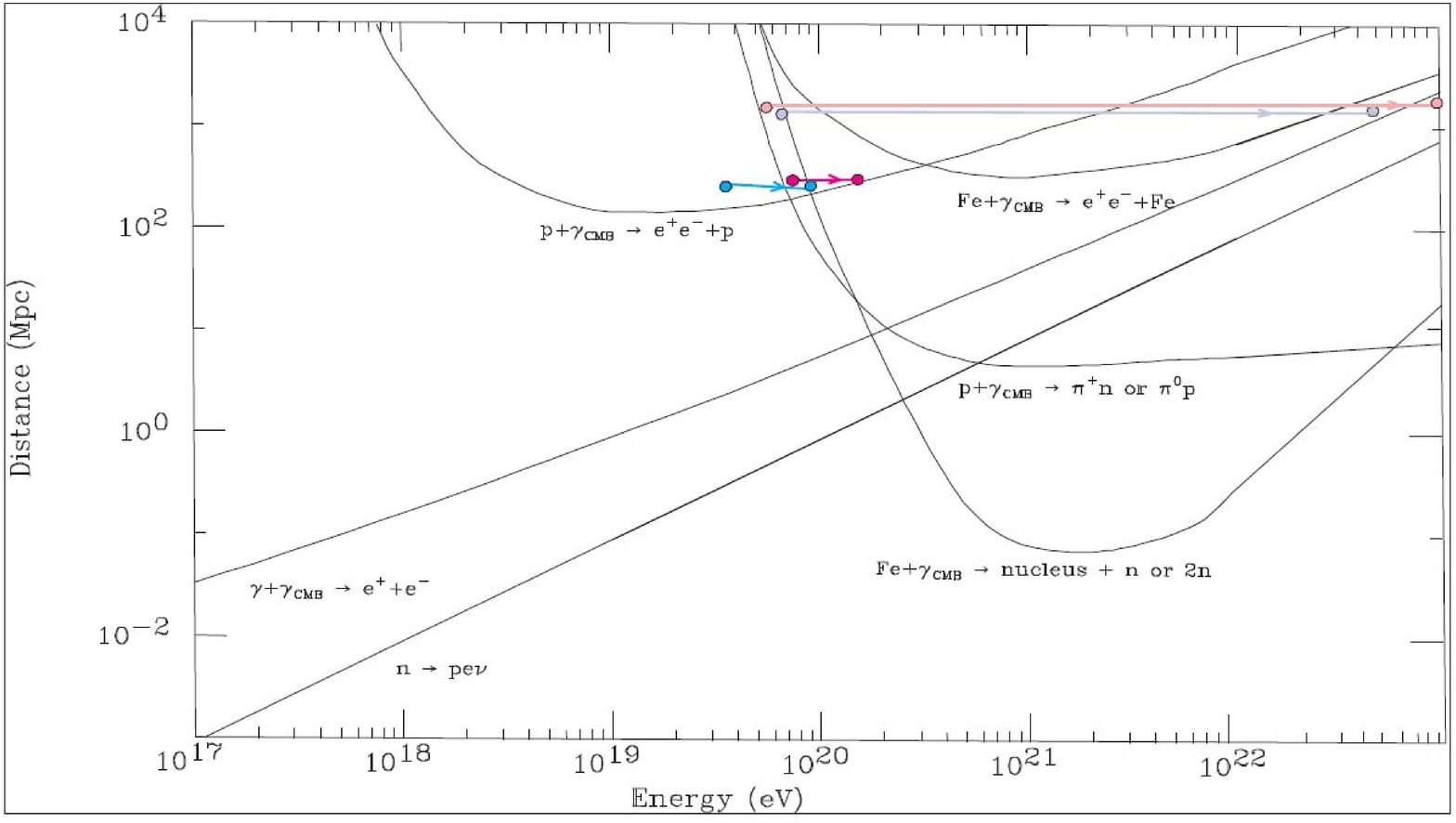}
\caption{\footnotesize{The nearly cosmic distances of most far
BL-Lac found in Gorbunov correlations (as the four shown in table
and in figure above) needs to arrive on Earth after consistent
energy losses; even in most optimistic scenarios their primary are
over Z-Burst energy edges \cite{Fargion04}.}} \label{fig5}
\end{figure}
\begin{figure}[t]
\includegraphics[width=65mm]{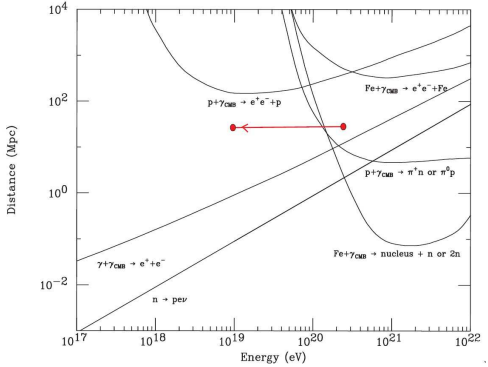}
\caption{\footnotesize{Z-Burst solution: read the text and
\cite{Fargion04}.}} \label{fig6}
\end{figure}
\begin{figure}[t]
\includegraphics[width=70mm]{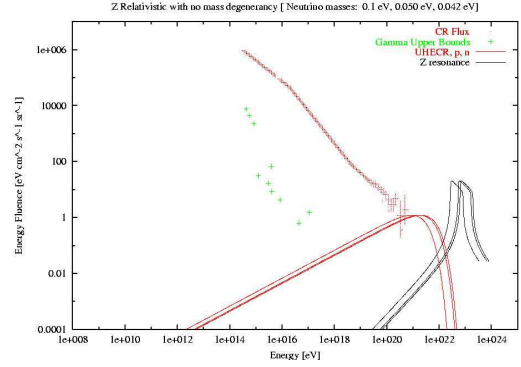}
\caption{\footnotesize{Each relic neutrino mass may interact at a
different incoming energy of UHE neutrino sources: the very recent
evidences for a very light neutrino mass may lead to the partial
breaking of the neutrino mass degeneracy as well as to the
splitting of the UHE resonance energy of Z-Burst Showering leading
to a possible future signature \textit{twin-towers} at UHECR bumps
in ZeV edges.}} \label{fig7}
\end{figure}
\begin{figure}[t]
\includegraphics[width=70mm]{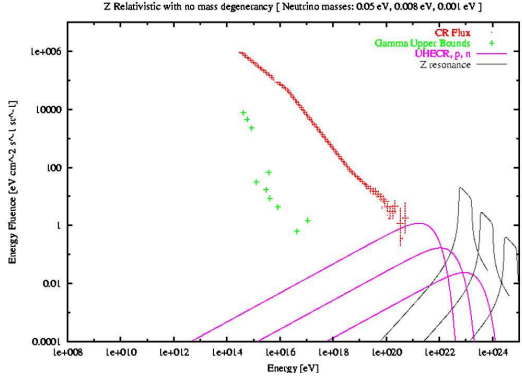}
\caption{\footnotesize{As above  relic neutrino mass may be very
light and totally non degenerated; each mass may  interact at a
different incoming neutrino energy: the lighter the $\nu$ mass the
less  is its clustering (Pauli principle) halo; the consequent
neutrino scattering  and its conversion probability in Z boson
 is lower and lower, as shown in corresponding fluence figure $tower$
height.}} \label{fig8}
\end{figure}
\begin{figure}[t]
\includegraphics[width=70mm]{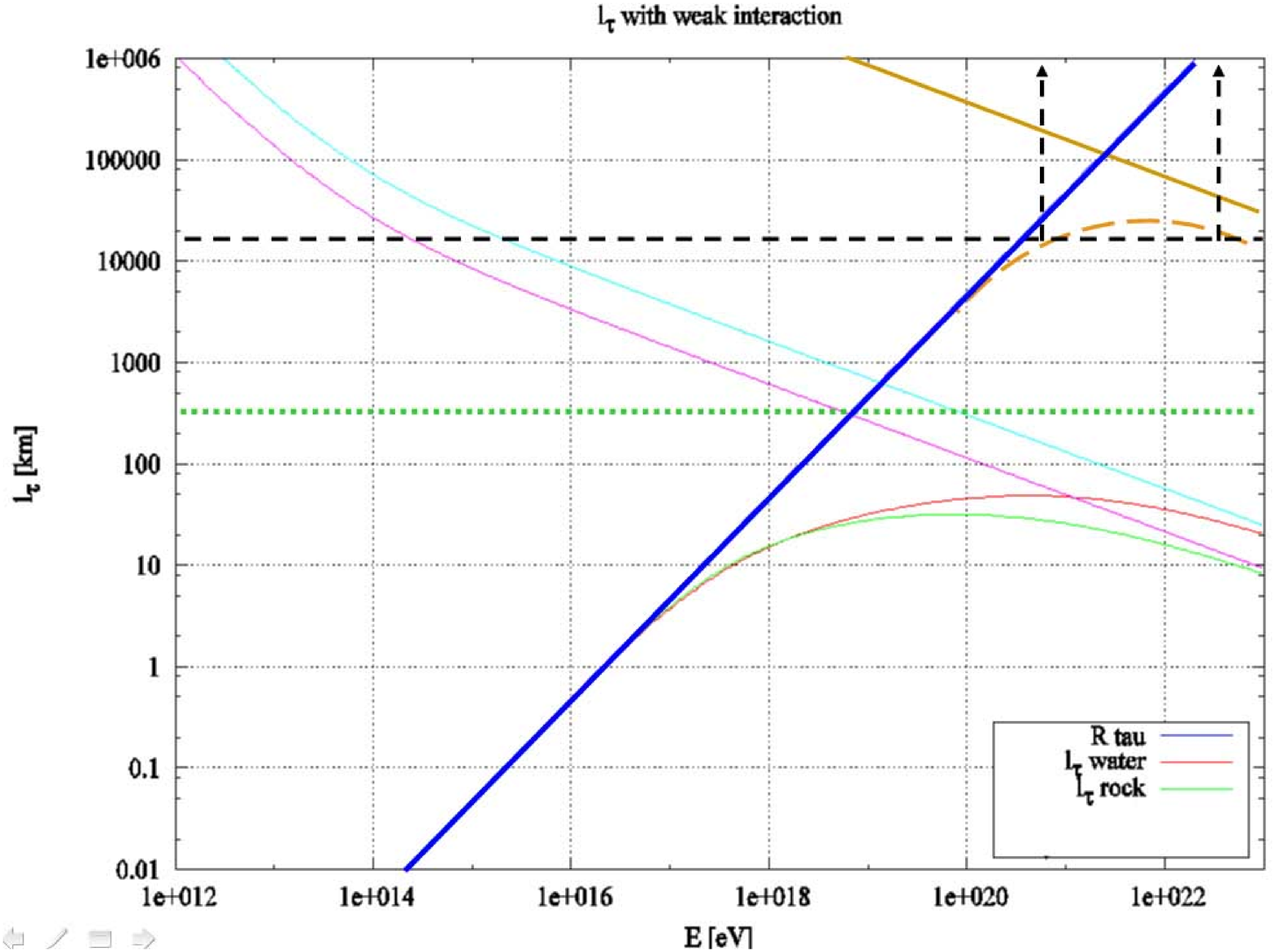}
\caption{\footnotesize{The interaction length of the tau and
parental UHE neutrino in water, rock and solar atmosphere. For the
best neutrino-tau conversion we considered here the sun limb where
the solar density and pressure are a fraction of the terrestrial
one (one third). The consequent approximated tau length is
crossing the solar cord at $2 \cdot 10^4$ km size, in
corresponding energies of ZeV values, shown by the two top
arrows.}} \label{fig9}
\end{figure}
The real Ultra High Energy (UHE)  Cosmic Rays (CR), UHECR, paradox
following K. Greisen, G.T. Zatsepin, V.A. Kuz'min, the famous GZK
cut off, lays not just in the (controversial but nevertheless
established) existence of UHECR spectra  above GZK ($\sim4\cdot
10^{19}$ eV) energies, but also in the extreme homogeneous and
isotropic map of their arrival directions. The absence of any
correlation or hint of nearby sources (G.C, Virgo, Super-Galactic
Maps) is, in our opinion, a major growing puzzle.
  Let us briefly summarize the UHECR state of art:
 \begin{enumerate}
  \item The eventual appearance of a  GZK cut-off as  claimed by HIRES
  experiment \cite{Hires06} is in partial disagreement to the earliest  AGASA
  data \cite{Takeda2003}. We note that a even more dramatic
discrepancy takes place \textit{all along} the UHECR spectra, by a
factor two or more, from EeV energy band energy up to GZK edges,
as shown in Fig.\ref{fig1}, \cite{Cronin2004}. This calibration
problem maybe the source of the overall disagreement. The AUGER
data seem to be consistent with HIRES more than AGASA, also at
lower energies.
  \item The absence of EeV correlation map, as claimed by AUGER,
with Galactic Center might signal the premature end of a timid but
exciting EeV neutron Astronomy, suggested by AGASA and SUGAR (see
Fig.\ref{fig2}).
  \item The absence of clear TeV correlation with
UHECR, as noted in \cite{Fargion04}, disfavors a direct proton
UHECR propagation at ten EeV energy \cite{Berezinsky 2002}, as
well as the very possible neutral nature of the final UHECR
associated to some BL-Lac source \cite{Hires06} (see
Fig.\ref{fig3}). A gamma ray precursor or afterglow at UHECR
arrival direction may disentangle the Z-Burst solution versus a
primary proton candidate \cite{Fargion04}.
  \item The Gorbunov, Tinyakov, Tkachev, and
Troitsky \cite{Gorbunov02} claim for an UHECR correlation with far
BL-LAcs has been somehow excluded (as it was), but nevertheless it
has been found out again, in a different set of records, by HIRES
itself \cite{Hires06}; therefore BL-Lac are still on-stage
candidate smoking     gun for UHECR. Their imprint might be found
in TeV gamma traces \cite{Fargion04}.
\end{enumerate}
The BL-Lac sources that might be the primary objects (by their UHE
jet emission toward us), are mostly located at distances quite
above GZK volumes (hundreds of Mpc). Therefore either one find a
reasonable way to transport these UHECR from cosmic edges to us
(with negligible or null energy losses) or one must advocate for
extreme random magnetic fields, $\sim 0.1\,\mu\,$Gauss, in
Universe to diffuse their maps. The latter seems quite
unrealistic. In a total different frame-work tens EeV protons,
long life and un-deflected in a nearly magnetic-free Universe has
been proposed recently \cite{Berezinsky 2002} (but see also
critical remarks in \cite{Fargion04}). Among the exotic solutions
for UHECR free arrival from cosmic edges, Lorenz invariance
violation has been more and more advocated (but also dismissed)
and even UHECR mirror neutrons have been postulated
\cite{Berezhiani06}. The Z-Burst model based on well established
neutrino masses (even at $0.1-0.05$ eV) and on possible UHE
neutrinos at ZeV energies \cite{Fargion-Mele-Salis99},
\cite{Weiler99}, \cite{Yoshida1998}, might solve quite naturally
the puzzle without new assumption on New Physics.
\begin{table*}[t]
\caption{BL Lacs and their real energies.}\label{tab:energiaalta}
\newcommand{\m}{\hphantom{$-$}}
\newcommand{\cc}[1]{\multicolumn{1}{c}{#1}}
\renewcommand{\tabcolsep}{1pc} 
\renewcommand{\arraystretch}{1.2} 
\begin{tabular}{cccccc}
\hline
    EGRET Name &  z & d (Mpc)&  $E_{obs} (10^{19}eV)$ & $E_{in} (10^{19}eV)$& Charge assignment \\
    \hline
    0808+5114 & 0.138 & 455 & 3.4 & 9.2 & 0 \\
    1052+5718 & 0.144 & 475 & 7.76 & 14.7 & 0,-1 \\
    1424+3734 & 0.564 & 1861 & 4.97 & $6\times10^{3}$ & 0,+1 \\
    1850+5903 & 0.53 & 1750 & 5.8 &$10^{4}$  & +1 \\
    \hline
    \end{tabular}
\end{table*}
\subsection{UHECR and the BL Lac connection.}
To make more compelling a UHECR - Z-Burst cosmic connection is the
very possible correlation between most distant BL Lacs (found
Gorbunov \cite{Gorbunov02} group as the ones shown in Table 1, or
in \cite{Hires06}) and UHECRs arrivals directions. These UHECR to
reach us with the observed energy and to overcome the electron
pairs energy losses, must be born at extreme energies, well above
the ones needed for Z-Burst models \cite{Fargion04}. Therefore the
Z-Burst UHE neutrinos are more realistic than other extreme
nucleon primary sources. It should be reminded that the very
recent data by HIRES and Gorbunov group are finding correlation in
a very narrow  UHECR arrival angle, pointing for a neutral UHECR
primary at ten EeV. Otherwise galactic field may slightly smear
their directions. One must notice that among the Z-boson decay
secondaries,  UHE $\gamma$ rays (by neutral pion at ten EeV
energy)  are required. Their presence, nevertheless have been
recently bounded by AUGER records \cite{Cronin2004}. However it
maybe be that just a $5-10\%$ of UHECR at ten EeVs energies are
enough to make the Hires BL-Lac connections with UHECR. This occur
at the same resonance energies of primary neutrino at ten ZeV band
and secondary nucleons at $10^{20}$ eV (see \cite{Fargion2001} and
\cite{Fargion03}). To make these arguments short and readable let
us summarize the fate of a final proton in a ten-EeV scenario
versus the Z-Burst model: in the first case in Fig.\ref{fig5}, the
primary UHECR, for most far BL Lac shown in Table 1, require
extreme primary energy. On the other side we see in Fig.\ref{fig6}
Z-burst secondary nucleon energy (after few tens Mpc flight) meets
the needed values.
\section{Z-Burst model for light neutrino mass.}
The idea to use the relic neutrinos as a beam dump where to
convert UHE cosmic $\nu$ and $\bar{\nu}$ hitting onto relic
$\bar{\nu}$ and $\nu$ is multi-faces: it is already tuned, with
present very possible light neutrino mass ($0.05-0.1$ eV) to UHE
$\nu$ at $E_{\nu}=4-2\cdot10^{22}$ eV and to the observed UHECR
edges above $10^{20}$ eV, (namely $8-4\cdot10^{20}$ eV), fitting
tens EeV photons secondaries. The light neutrino halo may extend
to tens Mpc offering a wide nearly isotropic cloud to capture
cosmic UHE ZeVs $\bar{\nu}$ and $\nu$. Fig.\ref{fig4} shows the
scenario of such a large halo where UHE ZeV $\nu$ shower to
observable  nucleons: $p,\;\bar{p}$, $n,\;\bar{n}$. The last
neutral ones  are few Mpc long life while the charged ones ar
confined in a GZK volume (ten Mpc). The second role in light
neutrino mass is the possibility to be so much light to split the
mass degeneracy into twin or triple values. Consequently the
Z-Burst may tune to different energies leading to different energy
injection at UHECR spectra edges. Present and future large array,
like Auger, may test this exciting UHE limits.
\subsection{UHECR  mirroring  neutrino masses.}
The different relic neutrino mass may be resonant in Z-boson
production at different UHE incoming neutrino. The highest energy
couple with the lightest neutrino mass. This has been noted and
explained \cite{Fargion2001}  and here summarized in
Fig.\ref{fig7} and in Fig.\ref{fig8}. The UHE incoming neutrino
\textit{meets} the lightest neutrino mass making a resonant
Z-boson at different energies; the highest incoming $\nu$ energy
scatters and resonates at  lightest $\nu$ masses. The clustering
of fermions are bounded by their $\nu$ masses (Pauli exclusion
principle): the lightest relic $\nu$ reach a lower density
contrast respect the heavier ones. Therefore the density and the
interaction probability may also be enhanced in multi Z-Burst peak
appearances at the UHECR edges (see Fig.\ref{fig7} and
Fig.\ref{fig8}).
\section{UHE $\nu_{\tau}$ skimming  the Sun.}
The same UHE neutrinos at ZeV energy, predicted in Z-Burst model,
might mix their primordial flavor, while travelling along stellar
or galactic distances; for present atmospheric $\nu$ mixing a
complete oscillation take place at a distance
$L_{\nu_{\mu}\rightarrow \nu_{\tau}}$:
 $$L_{\nu_{\mu}\rightarrow
\nu_{\tau}}=32\; pc\biggl(\frac{E_{\nu}}{10^{21}eV}\biggr)
  \biggl({\frac{\Delta m_{ij}^2}{2.5 \cdot 10^{-3}eV ^{2}}}\biggr)^{-1}
$$ giving life to the most rare $\tau$ lepton components.
These UHE $\nu_{\tau}$ (of Z-Burst nature), while interacting on
terrestrial crust maybe source of  skimming neutrinos (HorTaus
\cite{Fargion1999}, \cite{Fargion 2002a}, preferentially at lower
EeV energy band) whose ZeV tau decay in flight is too long,
$\simeq 5 \cdot 10^4$ km, to be contained in Earth air atmosphere.
It has been recently noted \cite{Fargion06} that wider atmosphere
layers are found in larger planets. The largest ones near Jupiter
and Saturn may reach a corresponding $\tau$ energy at $4-6 \cdot
10^{19}$ eV. At ZeV energies the widest and nearest atmosphere
horizons are found on our Sun.  However such ZeV
$\nu_{\tau}\rightarrow \tau$ solar grazing showers, while in
general being discontinue, are making brief millisecond flashes.
These showers may  manifest themselves in wide areas (on far Earth
detectors or satellites) by their overlapping and nearly
persistent bremsstrahlung secondary pairs which are  blazing into
X-ray flare. This occurs  while an observer is crossing into the
beam showers  cone radiated all along  the sun limb edges;  the
solar skimming bremsstrahlung shower may mimic weakest solar
flares or pulsed flashes within a narrow (few seconds) duration,
due to the sharp density profile size of the sun ($\simeq 200 km$)
where the phenomena occurs; these showers may be  correlated to
known sources arrival directions \cite{Fargion06}.
\subsection{UHE $\nu_{\tau}$ and $\tau$ interaction lengths.}
To estimate the UHE neutrino $\nu_{\tau}$ behavior while skimming
the solar atmosphere, making its UHE  $\tau$ whose later flight
and decay may blaze at solar  edges we show (see Fig.\ref{fig9})
their interaction lengths in respect to the terrestrial cases. The
tau showering at horizons has a long story \cite{Fargion1999},
\cite{Fargion 2002a}, \cite{Fargion2004}, \cite{Fargion2004b},
\cite{Fargion2005}, \cite{Bertou2002}, \cite{Cao}. The interaction
length play a key role: the $\tau$ energy losses in solar
atmosphere allows to reach distances as large as $2\cdot 10^{4}$
km, because of the much large solar radius (than Earth one, by a
factor $109$) and its larger height growth ($25$ times the
terrestrial ones); see also \cite{Fargion06}. The consequent place
where best tau skimming occur has an energy a ZeV edges, as shown
in Fig.\ref{fig9}.
\section{Conclusions}
Z-Burst is still an open solution to UHECR puzzling isotropy, BL
Lac Connections and possible future showering in solar edges.
Z-Burst may reflect itself, because of non-degenerated neutrino
mass splitting, into a surprising \textit{anti-GZK} bump
modulation in ZeV UHECR  spectra edges. The field is growing and
need to be carefully followed in view of the Auger near future
response.

\end{document}